\documentclass[aps,showpacs,twocolumn]{revtex4}
\usepackage{graphicx}
\usepackage{amsmath}
\usepackage{bm}

\begin{document}

\title{Dynamical entanglement and chaos: the case of Rydberg molecules}
\author{M.~Lombardi}
\author{A.~Matzkin}
\affiliation{Laboratoire de Spectrom\'{e}trie physique (CNRS Unit\'{e}
5588), Universit\'{e} Joseph-Fourier Grenoble-1, BP 87,
38402 Saint-Martin  d'H\`eres, France}

\begin{abstract}
A Rydberg molecule is composed of an outer electron that collides
on the residual ionic core. Typical states of Rydberg molecules
display entanglement between the outer electron and the core. In
this work we quantify the average entanglement of molecular
eigenstates and further investigate the time evolution of
entanglement production from initially unentangled states. The
results are contrasted with the underlying classical dynamics,
obtained from the semiclassical limit of the core-electron
collision. Our findings indicate that entanglement is not simply
correlated with the degree of classical chaos, but rather depends
on the specific phase-space features that give rise to inelastic
scattering. Hence mixed phase-space or even regular classical
dynamics can be associated with high entanglement generation.
\end{abstract}

\maketitle

\section{Introduction}

Simply stated, a Rydberg molecule is composed of a highly excited
electron orbiting around a compact ionic molecular core,
containing the nuclei and the tightly bound other electrons. Most
of the time, the outer electron is very far from the core. It is
spatially well-separated from it, and the core and electron
dynamics are uncoupled, the core's dynamics being characterized by
its rotational energy. However, the outer electron periodically
scatters on the molecular core. In quantum-mechanical terms the
electron and core dynamics get coupled and the collision induces
phase-shifts in the wavefunction of the outer electron, known as
quantum defects \cite{fano70,jungen96}. This process has a
well-defined classical counterpart
\cite{lombardi88}: the electron is kicked by the core, resulting
in a change of the electron's angular momentum relative to the
core by an angle which depends on the quantum defects. During the
kick the core and the outer electron may exchange energy, so that
in general the rotational state of the core has changed after the
collision. Although a molecule is intrinsically a quantum object,
several properties, such as the stroboscopic effect seen in laser
excitation or the statistics of the energy levels have been shown
to depend on the underlying classical dynamics
\cite{lombardi88,lombardi93}.

There is however a distinctive quantum feature that has no
classical counterpart, which is readily touched upon by observing
that a typical molecular state is described by quantum mechanics
as a superposition of the different core states available to the
molecule, each state being associated with the corresponding outer
electron. This superposition is the result of the electron-core
entanglement produced by the collision.

The main goal of this paper is to investigate the entanglement
dynamics and in particular its dependence on the underlying
classical regime. Indeed there has been a growing interest in recent
years to correlate the entanglement production of a quantum system
with the dynamics of the corresponding classical system. For example
the time-dependent entanglement production was investigated on the
$N$ atom Jaynes-Cummings model (a single mode field interacting with
a 2-level spin), where the initial product-state wave-packet was
chosen to lie in different regions of classical phase-space
\cite{furuya etal98}. Similar studies were undertaken for kicked
tops \cite{miller sarkar99,tanaka etal03,laksha04}. The initial
claims \cite{furuya etal98,miller sarkar99} by which entanglement
would systematically increase with chaos were revised in subsequent
works. One of the problems was to unequivocally define the classical
counterpart for such systems, which is far from being
straightforward for the spin-boson case and has been discussed for
the coupled kicks system. A semiclassical approach \cite{jacquod04}
based on the average properties of phase-space concluded that both
the coupling strength (in the form of classical correlators) and the
global classical dynamical regime were important in understanding
the entanglement rate. However the relevance of universal relations
relating the generation of entanglement to the dynamics of the
corresponding classical system are still being debated: the form of
the initial quantum state is known to play a role
\cite{kus04,seligman gorin03,fujisaki04}, and from a more general
viewpoint realistic systems usually display more complex dynamics
than simple systems that present a uniform behavior over all points
of phase-space. In this respect the choice of Rydberg molecules to
investigate entanglement generation may be fruitful: these real
systems are theoretically well described (by quantum defect theory),
the semiclassical limit is readily obtained, and the classical
dynamics is sufficiently simple to be well understood (at least
qualitatively) without being  too simplistic.

This work is organized as follows. In Sec.~II we briefly recall
some basics concerning the quantum theory of simple Rydberg
molecules and describe the classical counterpart of such systems,
insisting on the relevant Poincar\'{e} surfaces of section. In
Sec.~III we determine the degree of the electron-core
entanglement, inferred from the linear entropy of the reduced
density matrix for the outer electron. We will calculate some
simple statistics on groups of eigenstates corresponding to
different classical regimes. We will then follow the
time-evolution of the entanglement from an initially unentangled
product state. This will be done for different initial states and
various dynamical regimes. The results will be discussed in Sec.~%
IV. We will see that the classical dynamics is reflected in the
entanglement generation; but the global dynamical regime is less
important than the specific classical effects that are quantum
mechanically translated into superpositions. Our closing comments
will be given in Sec.~V.

\section{Rydberg molecules - quantum and classical}

\subsection{Quantum phase-shifts}

A Rydberg molecule is composed of two elements: a highly excited
electron -- the Rydberg electron, on the one hand, and the ionic
molecular core, positively charged, containing the nuclei and the
tightly bound other electrons on the other hand. The Rydberg
electron is usually far from the core and only senses the
long-range Coulomb field produced by the core, irrespective of the
complex interactions involving the core particles. The total
energy of the molecule $E$ is consequently partitioned between the
energies of the core and of the Rydberg electron.

For definiteness, we will take a diatomic molecule for which vibration can
be neglected: rotation is then the only motion available to the core. Since
the molecule is isolated, the total angular momentum $J$ and its projection
$M$ on an axis fixed in the laboratory are conserved. We have
\begin{equation}
\mathbf{J}=\mathbf{N}+\mathbf{L},
\end{equation}%
i.e. $\mathbf{J}$ results from the addition of the angular momenta
of the core, $\mathbf{N},$ and of the Rydberg electron
$\mathbf{L}$. We shall assume that $L$ is conserved, as is often
the case, and the standard addition of angular momenta gives
\begin{equation}
\left| J-L\right| \leq N\leq \left| J+L\right| .  \label{e10}
\end{equation}%
Each state of the core is therefore labeled by $\left|
N\right\rangle $. Recall that it also follows from the addition of
angular momenta that whereas $N$ and $L$ are well-defined, their
projections $M_{N}$ and $m$ on the reference axis are not, since
only the total projection $M=M_{N}+m$ is well-defined. Therefore
the notation $\left| N\right\rangle $ also contains the angular
state of the Rydberg electron, via the geometrical angular momenta
couplings. As is well known, the energy of the core $E^{+}$
depends
on $N$ through a rotational constant $B_r$,%
\begin{equation}
E_{N}^{+}=B_r N(N+1),
\end{equation}%
and the total energy of the molecule is thus%
\begin{equation}
E=B_r N(N+1)+\epsilon _{N}  \label{e12}
\end{equation}%
where $\epsilon _{N}$ is the energy and $\nu _{N}$ the effective quantum
number of the Rydberg electron,%
\begin{equation}
\epsilon _{N}=\frac{-1}{2\nu _{N}^{2}}
\end{equation}%
(atomic units are used throughout). Note that for a given value of
$E$, $\epsilon $ implicitly depends on the state of the core: the
electron is more or less excited depending on whether the core has
a large or small rotation number $N$. Here we will only deal with
bound states (i.e., $E$ is below the lowest ionization threshold).
The wavefunction corresponding to the Rydberg electron in the
Coulomb field of a
core in the rotational state $\left| N\right\rangle $ is thus%
\begin{equation}
\phi _{N}(E,r)=\left| N\right\rangle f_{L}(E-E_{N}^{+},r),  \label{e14}
\end{equation}%
$r$ being the radial coordinate and $f_{L}$ the Coulomb function regular at
the origin.

Now when the electron significantly approaches the core, the interaction
between the core particles and the electron cannot be neglected. These
interactions are embodied in the short-range potential $V$, that is
negligible beyond the core radius $r_{0}$. Thus the total Hamiltonian of the
molecule is
\begin{equation}
H=H_{0}+V,
\end{equation}%
with the functions $\phi _{N}$ being eigenstates of $H_{0}$. From
scattering theory, it is well-known that the wavefunctions for $H$
are obtained from

\begin{equation}
\psi _{N}(E,r)=\phi _{N}(E,r)-\sum_{N^{\prime }}g_{L}(E-E_{N^{\prime
}}^{+},r)\left| N^{\prime }\right\rangle K_{N^{\prime }N},  \label{e15}
\end{equation}%
where $g_{L}$ is the Coulomb function irregular at the origin and $K$ is the
scattering matrix; $K_{N^{\prime }N}$ gives the transition probability
amplitude between states $\phi _{N}$ and $\phi _{N^{\prime }}$ during the
collision. The eigenfunctions of the total Hamiltonian $H$ are obtained by
the superposition
\begin{equation}
\psi (E,r)=\sum_{N}Z_{N}(E)\psi _{N}(E,r).  \label{e19}
\end{equation}%
The coefficients $Z_{N}(E)$ are obtained from quantum defect
theory \cite{jungen96} by imposing the appropriate boundary
conditions at infinity, yielding also the discrete eigenvalues
$E$. We thus see from Eqs.~(\ref{e15}) and (\ref{e19}) that a
generic wavefunction for a molecular Rydberg state involves a
superposition of core states with different rotational numbers.
This is caused by the short-range potential $V$ that transforms
the \emph{product state} $\phi _{N}$ [Eq.~(\ref{e14})] into the
\emph{entangled state} $\psi $. In practical computations,
Eq.~(\ref{e19}) is of awkward use because taken individually each
of the functions $\psi _{N}(E,r)$
diverges radially. Eq.~(\ref{e19}) is therefore rewritten as%
\begin{equation}
\psi (E,r)=\sum_{N}B_{N}(E)F_{L}(\epsilon _{N},r)\left| N\right\rangle
\label{e20}
\end{equation}%
where the $F_{L}(\epsilon _{N},r)$ are the effective channel functions,%
\begin{equation}
F_{L}(\epsilon _{N},r)\equiv \sin \beta (\epsilon _{N})f_{L}(\epsilon
_{N},r)-\cos \beta (\epsilon _{N})g_{L}(\epsilon _{N},r),
\label{e20b}
\end{equation}%
which by construction converge as $r\rightarrow \infty $. $\beta
(\epsilon _{N})\equiv \pi (\nu _{N}-L)$ is precisely the total
phase accumulated at $r\rightarrow \infty $. The coefficients
$B_{N}$ are obtained from the $Z_{N}$ by matching Eqs.~(\ref{e19})
and (\ref{e20}).

An additional subtlety arises from the use of frame transformations: the
wavefunctions given above were described in the laboratory frame. However
when the electron collides on the core, a description in the molecular
frame, attached to the core rotation, is more appropriate, because in the
core region the Rydberg electron senses the cylindrical field aligned along
the molecular axis. Thus only the projection of $\mathbf{L}$ on that axis,
traditionally denoted by $\Lambda$, matters when describing the collision:
the phase-shifts induced by the collision on the Rydberg electron's
wavefunction are known as quantum defects and denoted by $\mu_{\Lambda}$.
The collision matrix $K$ appearing in Eq.~(\ref{e15}) is obtained by
expressing the phase-shifts $\mu_{\Lambda}$  in the laboratory frame %
\cite{fano70}.

\begin{figure}[tb]
\includegraphics[angle=270,width=2.0in]{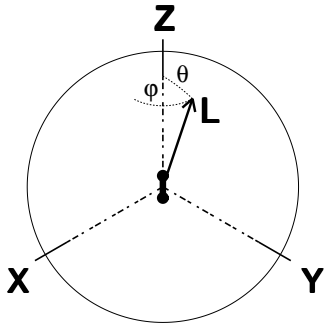}
 \caption{Molecular reference frame. The core axis is along $OZ$ and its
angular momentum $\mathbf{N}$ is along $OX$}%
\label{figframe}
\end{figure}

\subsection{Classical kicks\label{ck}}

The classical counterpart of the quantum model introduced above is the
following \cite{lombardi88}. When the Rydberg electron is far from the
core, it follows a pure Coulomb (Kepler) orbit with an angular momentum $%
\mathbf{L}$ fixed in space. Meanwhile the molecular core rotates
freely with a rotational energy $N(N+1)/2I$ depending on the
core's angular momentum $N$ and moment of inertia $I=1/2B_r$. Seen
in the molecular frame, $\mathbf{L}$ precesses around
$\mathbf{N}$, i.e. it turns around the $\mathbf{N}$ axis with a
constant angle (see Fig.~\ref{figframe}). Now when the outer electron
approaches the core, it gets kicked by the molecular axis. This
kick results in a change in the direction of $\mathbf{L}.$
$\mathbf{N}$ adjusts accordingly, since the total angular momentum
$\mathbf{J}$ is conserved. To visualize the effects of the kick,
it is therefore sufficient to follow the evolution of $\mathbf{L}$
in the molecular frame.

Since during the collision the Rydberg electron feels the
cylindrical field
due to the molecular axis, $\theta $ cannot change and thus only the angle $%
\varphi $ varies (see Fig.~\ref{figframe}). This variation, denoted $\delta
\varphi $, is the deflection angle of the plane of the classical
Kepler orbit. The relation between the classical deflection angle
and the quantum scattering phase-shifts is well known from the
semiclassical approximation to the scattering amplitude \cite{ford
wheeler59}. Here it takes the form
\begin{equation}
\delta \varphi =2\pi \frac{\partial \mu _{\Lambda }}{\partial \Lambda },
\label{e21}
\end{equation}%
i.e. the strength of the kick, reflected in the amplitude of the deflection
angle, is the classical counterpart of the dependence of the phase-shifts on
$\Lambda $. The precise form of this dependence depends on the particular
molecule at hand. However, for a typical molecule, the dependence can be
taken in the form%
\begin{equation}
\mu _{\Lambda }=\mu _{0}-k\frac{\Lambda ^{2}}{4\pi L}.  \label{e23}
\end{equation}%
Classically the coupling parameter $k$ gives the strength of the kick since
it follows from Eqs.~(\ref{e21}) and (\ref{e23}) that%
\begin{equation}
\left| \delta \varphi \right| =k\frac{\Lambda }{L}=k\cos \theta .
\end{equation}

\begin{figure}[tb]
\includegraphics[width=0.9\columnwidth]{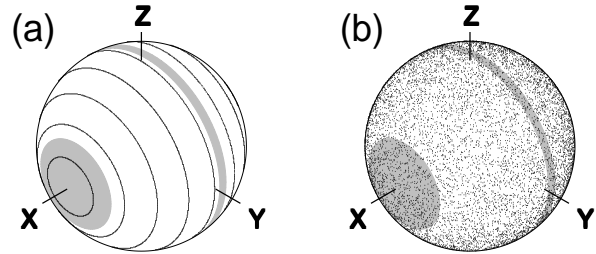}
\caption{Two limiting Poincar\'e surfaces of sections. (a) no coupling $k=0$
(b) large coupling $k=10$ (in the generic case, see Fig.~\ref{figPSS} below).
The shaded areas correspond to quantum states with minimal
($J-L=40$), and mean ($J=50$) values of $N$ (see Sec.~III). Their
widths correspond to $\Delta N=1$. The mean $N=J$ case is not on
the equatorial $OYZ$ plane, as can be understood by squaring
$\mathbf{J}=\mathbf{L}+\mathbf{N}$ (where $\mathbf{N}$ is
``horizontal'', parallel to $OX$) when the lengths of $N$
and $J$ are equal.}%
\label{figPSS0max}
\end{figure}

To visualize the classical dynamics, a Poincar\'{e} surface of
section is obtained by plotting the position of $\mathbf{L}$ after
each kick. This is most naturally done in the molecular frame where
the position of $\mathbf{L}$ is given by the polar angles
$(\theta,\varphi)$, as shown in Fig.~\ref{figframe}, with $%
\varphi$ being canonically conjugate to $L_{z}$. This is the
molecular reference frame used in quantum mechanics. However, as
in our previous works \cite{lombardi88,lombardi93,lombardi04}, we
prefer to add an extra rotation around $OZ$ to bring the $OX$ axis
along $\mathbf{N}$. This last rotation is not canonical but the
classical motion is seen more naturally in this frame (see the
discussion in Ref.~\cite{lombardi04}). The position of
$\mathbf{L}$ is thus plotted after each kick, when the electron
comes out of the core. Of course, since $\mathbf{J=L+N}$ is
conserved, following the position of $\mathbf{L}$ is tantamount to
knowing the fate of $\mathbf{N}$. Two extreme examples of
Poincar\'{e} sections are given in Fig.~\ref{figPSS0max}. In the first case,
Fig.~\ref{figPSS0max}(a), the positions of $\mathbf{L}$ follow circles around
$\mathbf{N}$ (the $X$ axis) giving an overall regular surface of
section. This means that on a given circle $\mathbf{L}$ is fixed
in space, as is always $\mathbf{J}$, and thus that $N=|\mathbf{J}-\mathbf{L}|$
is constant. In the second case, Fig.~\ref{figPSS0max}(b), the surface of
section is clearly characteristic of a chaotic phase space and no structure
arises by following the successive positions of $\mathbf{L}$ after each kick.

\section{Entanglement dynamics}

\subsection{General Remarks}

Generically, in a Rydberg molecule the core and the outer electron
are entangled; in the situation examined in this work, the core is
in a superposition of different rotational states. Each rotational
state is defined by a given value of $N$,
and to each core state is associated an outer electron with an energy
$\epsilon (N)$ given by Eq.~(\ref{e12}). Hence the outer electron
is in a superposition of different channels which are
distinguished in the present model by a different energy,
depending on $N$. Classically of course there is no superposition
in $N$: the rotational energy of the core, and thus the energy of
the outer electron, is at each instant unique and well defined. A
change in $N$ can only be the dynamical result of an inelastic
kick. Note
that classically as well as quantum mechanically, the angular momenta
$\mathbf{L}$ and $\mathbf{N}$ are coupled, since $\mathbf{J}$ is
conserved. As is well-known, the composition of angular momenta in
quantum mechanics results in couplings due to the fact the
projections of the angular momenta vectors cannot be
simultaneously defined. This entanglement of geometrical nature
(as in EPR pairs) should not be confused with the dynamical
entanglement generated by the potential interaction between the
outer electron and the core. Geometrical entanglements due to
angular momenta coupling do not play any role in this work, as we
are only interested in the dynamical one which only depends on
$\left| \mathbf{N}\right| $. This implies that a product state is
given by the channel functions $\phi _{N}(E,r),$ defined by Eq.~%
(\ref{e14}), i.e. only the \emph{radial} coordinate of the outer
electron is separable from the core whereas its orbital angular
momentum is necessarily geometrically coupled to the core's
angular momentum $\mathbf{N}$ (since $\mathbf{J}$ is conserved and
$H_{0}$ has spherical symmetry). Hence when referring to partial
traces on the outer electron we will mean a trace over its sole
radial coordinate and conversely a partial trace over the core
includes tracing over the angular coordinates of the electron.

To quantify entanglement we will determine the linear entropy
$S_{2}$ associated with the reduced density matrix $\rho _{e}$
descibing the outer electron,
\begin{equation}
\rho _{e}=\mathrm{Tr}_{c}\rho =\sum_{N}\left\langle N\right| \rho \left|
N\right\rangle ,  \label{e30}
\end{equation}%
where $\rho \equiv \left| \psi \right\rangle \left\langle \psi \right| $ is
the density matrix of the system and $\mathrm{Tr}_{c}$ (resp. $\mathrm{Tr}%
_{e}$) refers to averaging over the core (resp. outer electron) degrees of
freedom. The reduced linear entropy is then defined by%
\begin{equation}
S_{2}=1-\mathrm{Tr}_{e}\rho _{e}^{2}.  \label{e31}
\end{equation}%
Strictly speaking, $S_{2}$ measures the degree of mixedness:
$S_{2}$ vanishes for a pure state and is maximum for a uniformly
mixed state. However when $S_{2}$ is associated with improper
mixtures in bipartite systems, it reflects the degree of
entanglement \cite{espagnat} and the linear entropy or equivalent
quantities such as the purity have routinely been employed as such
\cite{furuya etal98,tanaka etal03,%
seligman gorin03,kus04,jacquod04,prosen05}. We will undertake two
different studies. First we will investigate the degree of
entanglement on stationary states and its dependence on the
collision phase-shifts, whose classical counterpart gives rise to
different dynamical regimes. This involves the computation of
simple statistics of $S_{2}(E)$ in an energy range for which the
classical dynamics does not vary. We will then investigate the
time evolution $S_{2}(t)$ from an initially (at $t=0$) product
state. This involves the determination of wavepacket dynamics. The
initial wavepacket can be made to lie and then evolve in zones
corresponding to different classical dynamics.

We will obtain numerical results for the following choice of
parameters: $J=50,$ $L=10$ yielding by Eq.~(\ref{e10}) 21 values
for $N$. Since Kronig's parity, i.e. parity by reflection at a
plane through the internuclear axis $OZ$ (which for fixed $L$ and
$J$ depends on the parity of $N$), is conserved, entanglement only
takes place between states of the same Kronig's parity. States of
$(+)$ and $(-)$ Kronig's parity, called positive and negative
states in standard spectroscopic notation \cite{HerzbergI50},
behave in the same way, and we will restrict our study to $(+)$
parity states; then $N$ can only take even values, so that a
typical state contains superpositions involving up to $11$ values
of $N$. These angular momenta numbers are considerably higher than
for typical diatomic molecules (they would better correspond to
models of large molecular compound Rydberg states). However higher
quantum numbers allow a finer comparison between quantum and
classical dynamics, without qualitatively affecting the
correspondence between them: see \cite{lombardi matzkin06}, where
partial results with angular quantum numbers typical of diatomic
molecules such as Na$_{2}$ were obtained
\footnote{In fact relative to \cite{lombardi matzkin06} we choose in
this work a situation with identical classical parameters (in particular the
coupling constant $k$ and the crucial ratio $T_e/T_c$), and thus
identical Poincar\'e surfaces of sections, but with an effective
$\hbar$ divided by 5, and thus all quantum numbers ($L, J, N,
\nu$) are multiplied by 5, and $B_r$ is divided by $5^4$.}.

\subsection{Degree of entanglement of eigenstates}

The eigenstates $\psi (E)$ of a Rydberg molecule are given by
[Eqs.~(\ref{e20},\ref{e20b})]
\begin{widetext}
\begin{equation}
\psi (E)=\sum_{N}B_{N}(E) \left( \sin \beta (E-E_{N}^+)f_{L}(E-E_{N}^+,r)-\cos \beta (E-E_{N}^+)g_{L}(E-E_{N}^+,r) \right) \left| N\right\rangle.
\end{equation}%
\end{widetext}
To quantify the degree of entanglement of a given eigenstate, we
compute $S_{2}(E)$. Variations from individual eigenstates are
smoothed out by calculating simple statistics for the bunch of
eigenstates sitting in an interval $\Delta E$. The requirement on
$\Delta E$ is that the classical dynamics does not change
appreciably within this interval. Several computations are performed
for different values of the coupling constant $k$. Each value of $k$
corresponds to a different collision matrix $K$ via Eq.~(\ref{e23}).

\begin{figure}[tb]
\includegraphics[height=1.5in,width=2.0in]{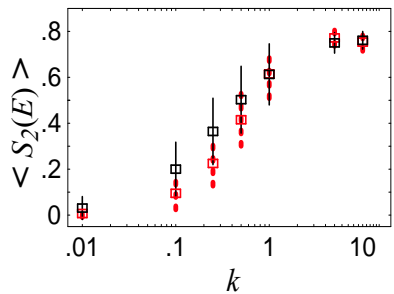}
 \caption{Average linear
entropy for generic (red boxes, color online) and resonant
$T_{e}=T_{0}$ states (black boxes), shown for different values of
$k$. The rms is also shown as red dashed (generic) and solid
(resonant) error bars.
}%
\label{fig3}
\end{figure}

\begin{figure}[tb]
\includegraphics[width=0.9\columnwidth]{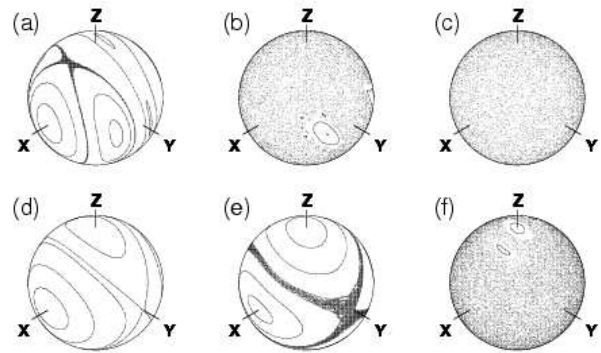}
 \caption{Poincar\'e surfaces of section for different values of the
 coupling constant $k$.
 Top: generic; bottom: resonant cases.
 (a,d) $k=0.25$. (b,e) $k=1$ (c,f) $k=10$.}%
\label{figPSS}
\end{figure}

The results are shown in Fig.~\ref{fig3}. The average and rms of $S_{2}$
are given for different values of $k$. Fig.~\ref{fig3} shows both the
results for a 'generic' situation which would be obtained for an
arbitrary choice of $E$ and the results for resonant eigenstates:
in the latter case $E$ is chosen such that the period of the
Rydberg electron $T_{e}$ is an integer multiple of half the period
of the core $T_{c}$. Classically, this corresponds to a situation
in which the electron sees the core in the same position on its
return as when it left the core region. Resonances  affect the
classical dynamics, essentially by retarding the appearance of
chaos. This is portrayed in Fig.~\ref{figPSS}, which shows Poincar\'{e}
surface of sections for several values of the coupling $k$ in both
the generic and the resonant cases. In the former case, chaos
appears even for a small value of $k$, whereas in the latter
configuration, chaos becomes significant for larger values the
coupling, and even for such a large value as $k=10$ an island of
regularity around the fixed point on the $Z$ axis is still
visible.

Figure~\ref{fig3} indicates that on average the degree of entanglement, as
measured by $<S_{2}>$ increases with $k$. However this does not
mean that entanglement is correlated with classical chaos. Thus
for example for $k=0.25$ the average entanglement is significantly
higher in the resonant case than in the generic case, despite
phase space being slightly more regular (see Figs.~\ref{figPSS}(a) and (d)).
For $k=1$ $<S_{2}>$ has the same value in the generic and resonant
cases although the dynamical regimes, as reflected in the surfaces
of section Fig.~\ref{figPSS}(b) and (e), are quite different.

\begin{figure}[tb]
\includegraphics[width=0.9\columnwidth]{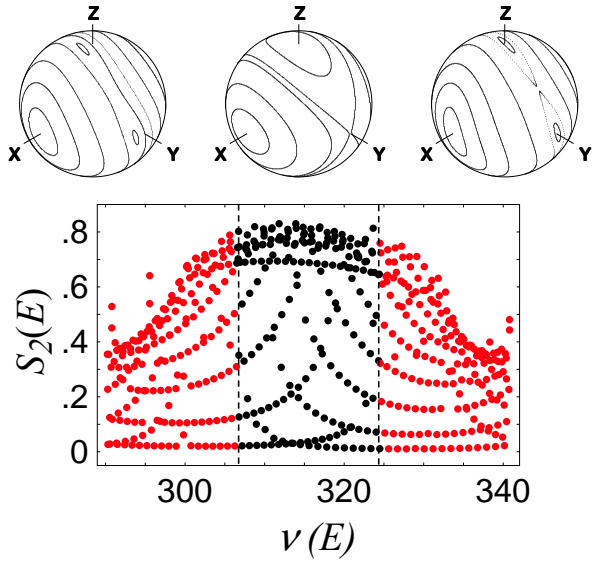}
\caption{Classical phase-space and linear entropy in the $k=0.5$ resonant
situation. (Top) Poincar\'e surfaces of sections at the left,
middle, and right ends of the energy interval shown at the bottom.
(Bottom) Linear entropy of the individual eigenstates. The black
dots between the dashed lines correspond to the states that
entered the statistics shown in Fig.~\ref{fig3}. Each eigenstate of energy
$E$ is labeled by $\nu(E)$, the
principal quantum number in the $N=J$ channel.}%
\label{figS2PS}
\end{figure}

We further illustrate the effect of a resonance both on the
classical dynamics and on the linear entropy for the case $k=0.5$.
Indeed, as the energy is appreciably moved away from the exact
resonance energy, the classical dynamics accordingly changes,
going back towards a generic situation. This is shown in
Fig.~\ref{figS2PS}~(Top): the Poincar\'{e} surface of section at the center is
plotted at the exact resonance energy. However as the energy
changes appreciably, the periods of the Kepler orbit and the core
rotation are significantly altered, suppressing the resonance.
Classically, the structure of phase-space is modified, as can be
directly seen on the surfaces of section plotted at both ends of
the energy range. At all energies phase-space is regular, but the
separatrices characterizing the resonance give way to island
chains and curves organized around $OX$. Fig.~\ref{figS2PS}~(Bottom) shows the
linear entropy for the \emph{individual} states lying within this
energy range. It may be seen that as we move away from resonance
the behavior of the linear entropy drops dramatically: most states
show a lower degree of entanglement. The degree of entanglement
clearly appears to be correlated with the changes in phase-space
induced by the resonance. The findings presented in Figs.~\ref{fig3}
and \ref{figS2PS}
will be discussed in Sec.~IV, but we may note that these results
indicate that entanglement is sensitive to the details of
classical phase space, not only to the global dynamical regime.
Note that the statistics for $k=0.5$ shown in Fig.~\ref{fig3} were done
from the 200 individual states lying within the dashed lines in
Fig~\ref{figS2PS}.

\begin{figure}[tb]
\includegraphics[width=0.9\columnwidth]{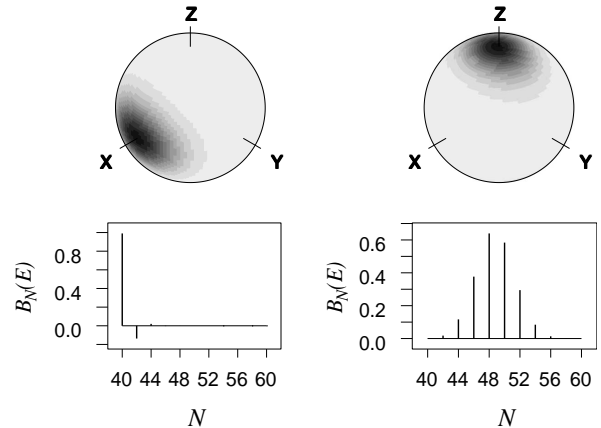}
\caption{Husimi plots (Top) and $B_N(E)$ coefficients (Bottom) for
two particular wavefunctions in the $k=0.5$ resonant situation near
the center of the interval plotted in Fig.~\protect\ref{figS2PS}.
Left level, with $\nu(E)=315.4233$, is nearly quantized on the
minimum value of $N=40$ as shown by the $B_N$ distribution. Compare
it's Husimi plot with shaded areas in Fig.~\ref{figPSS0max}. It has
$S_2(E)=0.03658$. The right level, with $\nu_E=315.6118$, quantized
nearby the $+OZ$ axis, spans a greater number of $B_N$ values. It's
Husimi plot overlaps with a greater number of $N$ circles indicated
in Fig.~\ref{figPSS0max}. It has $S_2(E)=0.6881$.} \label{figHusiBN}
\end{figure}

To grasp the relationship between quantum entanglement and the
classical dynamics for individual eigenstates, we project the
wavefunctions in mock phase-space on the surface of section by
drawing the Husimi plots for two  eigenstates having different
values of $S_2$. We give in Fig.~\ref{figHusiBN} the Husimi plot
along with the 11 $B_N$ coefficients of two particular wavefunctions
lying very near the center of the resonance shown in Fig.
\ref{figS2PS}. The first one is (accidentally) a nearly pure $N=40$
wavefunction. Its Husimi plot shows that it is quantized near the
$+OX$ axis, as expected from the gray zone displayed in
Fig.~\ref{figPSS0max}. Being a nearly pure product, it has thus a
very low linear entropy, $S_2=0.03658$. The second wavefunction is
(also accidentally) quantized nearly on the $+OZ$ axis. Its phase
space extension is nearly the same, but the decomposition on the
$B_N$ basis spans more values of $N$, as shown in the lower part of
the figure. This is understood by considering the overlap of its
Husimi plot with $OX$-centered circles, each such circle
corresponding to a value of $N$. Its linear entropy is thus
correspondingly much higher: $S_2=0.6881$.

\subsection{Dynamical evolution of entanglement}

We consider the time dependence of the generation of entanglement
from an initial product state. Assume the system has been prepared
so that at $t=0 $ the core has a well defined rotational state
$\left| N_{0}\right\rangle $ whereas the outer electron is
radially localized at the outer turning point of the Kepler orbit,
several thousand atomic units away from the core. The wavepacket
attracted by the Coulomb interaction moves towards the core and
collides on it at $t\approx T_{e}/2$. The collision results in
entanglement, since the outgoing waves of the Rydberg electron are
in a state of superposition, each outgoing channel being attached
to a core in a different quantum state. Subsequent collisions with
the core result in further entanglement, whereas the spreading of
the radial wavepacket quickly results in a continuous
core-electron interaction.

We take the initial state to be%
\begin{equation}
\psi (t=0,r)=F_{loc}(r\approx r_{tp})\otimes \left|
N_{0}\right\rangle \label{e31b}
\end{equation}%
where the radial function is
\begin{equation}
F_{loc}(r)=\sum_{n}e^{-\left[ (n-n_{0})/2\Delta n\right]
^{2}}R_{nL}(r)
\end{equation}%
with an appropriate normalization factor. $R_{nL}(r)$ are the
standard radial functions of the hydrogen atom (regular Coulomb
functions) and the Gaussian form of the coefficients are known to
ensure localization \cite{nauenberg90}. $n_{0}$ is chosen so that
the central component of the wavepacket matches the energy of the
corresponding classical regime under
study. At later times the wavefunction is given by%
\begin{equation}
\psi (t,r)=\sum_{E}\sum_{N}\mathcal{B}_{N}(E)e^{-iEt}F_{L}(E-E_{N}^{+},r)
\left| N\right\rangle ,  \label{e32}
\end{equation}%
where%
\begin{equation}
\mathcal{B}_{N}(E)=B_{N}(E)B_{N_{0}}(E)\left\langle F_{L}(E-E_{N}^{+})
\right| \left. F_{loc}\right\rangle  \label{e34}
\end{equation}%
where the coefficients $B_{N}(E)$ and effective channel radial functions $%
F_{L}(E-E_{N}^{+})$ were given above (cf Eqs.~(\ref{e20}) and
(\ref{e20b})). The radial overlap $\left\langle F_{L}(\epsilon
_{N})\right| \left. F_{loc}\right\rangle $ is determined
analytically as a particular instance of the scalar product
$\left\langle F_{L}(\epsilon )\right| \left. F_{L}(\epsilon
^{\prime })\right\rangle $ given by \cite{bell seaton}
\begin{equation} \label{sxsx}
\left\langle F_{L}(\epsilon)\right|  \left. F_{L}(\epsilon^{\prime
})\right\rangle
=\frac{\sin\pi(\nu-\nu^{\prime})}{\pi(\nu-\nu^{\prime})}
\end{equation}
multiplied by the relevant normalization factors \footnote{This must
be done with care, as Eq. (\ref{sxsx}) supposes a channel
normalisation to unity, whereas the  MQDT wavefunctions are
normalized globally per unit energy \cite{jungen96,lombardi04}.
Failure to normalize correctly the radial channel functions $F_{L}$
 results in an imperfect cancellation when taking the scalar product of two different
 eigenstates, leading to errors of the order of several $10^{-3}$. Still perfect
cancellation does not occur, as errors due to the approximate nature
of MQDT itself remain. These errors are however in the fully
negligible, in the range $10^{-5}-10^{-6}$, as we have checked by an
independent test calculation which makes an "exact" computation with
a given short range potential.}.

In principle, the computation of the linear entropy associated with the
reduced density matrix is straightforward. From%

\begin{equation}
\rho (t)=\left| \psi (t)\right\rangle \left\langle \psi (t)\right|
\end{equation}%
we obtain the purity $\mathrm{Tr}_{e}\rho _{e}^{2}(t)$ and the
reduced linear entropy $S_{2}(t)$ as
\begin{widetext}
\begin{eqnarray}
&&\mathrm{Tr}_{e}\rho _{e}^{2}(t)=\int \left\langle r\right| \rho
_{e}^{2}\left| r\right\rangle r^{2}dr  \label{e36} \\
&&\mathrm{Tr}_{e}\rho _{e}^{2}(t)=\sum_{NN^{\prime }}\left| \sum_{EE^{\prime
}}e^{-i(E-E^{\prime })t}\mathcal{B} _{N}(E)\mathcal{B} _{N^{\prime}}
(E^{\prime})\left\langle F_{L}(E^{\prime }-E_{N^{\prime }}^{+}\right| \left.
F_{L}(E-E_{N}^{+})\right\rangle \right| ^{2}.  \label{e37}
\end{eqnarray}%
\end{widetext}
The radial closure relation needs to be introduced in Eq.~(\ref{e36}) given
that the effective radial functions that play the role of the basis are
overcomplete.

\begin{figure}[tb]
\includegraphics[height=2.3in,width=2.15in]{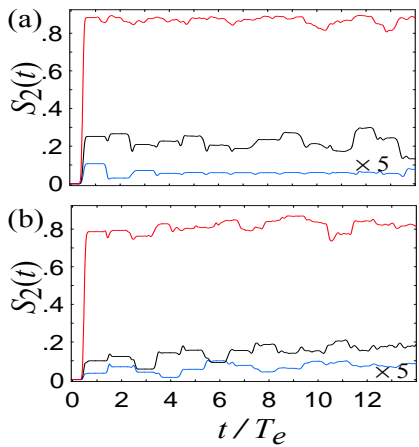}
\caption{Short time variation of the linear entropy generated from
the initial product state $F_{loc}(r\approx r_{tp})\otimes \left|
N_{0}=40\right\rangle$ for different collision matrices labeled by
the value of the coupling strength $k$: from top to bottom $k=10$
(color online, red), 1 (black) and 0.25 (blue). $t$ is given in
units of the Kepler period $T_{e}$. (a) The parameters are chosen
so that the corresponding classical dynamics falls in the
\emph{generic} phase-space case. (b) Same as (a) for the
\emph{resonant} case $T_{e}=T_{0}$. In both (a) and (b) the
$k=0.25$ curve is multiplied by a factor 5. }%
\label{short-time-1}%
\end{figure}

\subsubsection{Short-time evolution}

We examine first the short time-evolution for three different
couplings (kick strength) -- $k=0.25$, $k=1$ and $k=10$ -- and two
different initial states distinguished by the value of $N_{0}$:
$N_{0}=J-L$ (that is the minimal value $N_{0}$ can take, cf Eq.~%
(\ref{e10})) and $N_{0}=J$. The corresponding zones in the surfaces
of section are shaded in Fig.~\ref{figPSS0max}. Recall that on the
surface of section a fixed value of $N_{0}$ corresponds to a line
circle around the $X$ axis, whereas a quantum state in mock
phase-space projected on the surface of section has a certain width,
as seen in Fig. \ref{figHusiBN}. The width of the initial state can
be roughly estimated by plotting the zone going from $N-0.5$ to
$N+0.5$, cutting the sphere approximately into the number of values
$N$ can take (here 11).

\begin{figure}[tb]
\includegraphics[height=2.3in,width=2.15in]{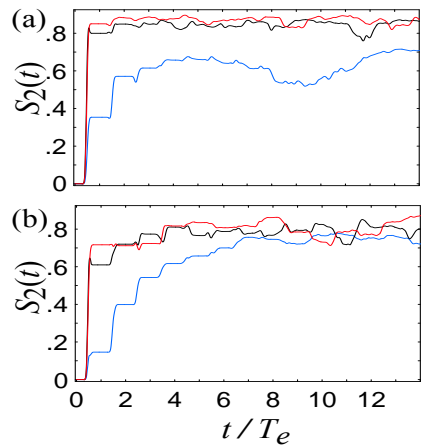}
\caption{Same as Fig.~\ref{short-time-1} but with the initial
product state given by $F_{loc}(r\approx r_{tp})\otimes \left|
N_{0}=50\right\rangle$ (the
$k=0.25$ curve is \emph{not} multiplied by a factor 5). }%
\label{short-time-2}%
\end{figure}

Fig.~\ref{short-time-1}(a) and (b) show the linear entropy as a function of
time (in units of the Kepler period) for the generic and resonant cases when
the initial state (\ref{e31b}) is chosen with  $N_{0}=J-L=40$.
Figure~\ref{short-time-2} gives the linear entropy
when the initial state is taken with $N_{0}=50$. In
Fig.~\ref{short-time-1}, we observe that in both
the generic and resonant cases $%
S_2(t)=1-\mathrm{Tr}_{e}\rho _{e}^{2}(t)$ increases with $k$:
entanglement is produced more rapidly and saturates at a higher
value. Comparing with the classical dynamics, we see that in the
generic case, the rise in entanglement generation accompanies the
classical transition to chaos. This remains true to a certain
extent in the resonant case, as mixed phase-space turns
progressively chaotic.

However when the initial rotational state lies near the center of
the surface of section (Fig.~\ref{figPSS0max}), the linear entropy takes
very high values irrespective of the classical dynamical regime. This
is particularly spectacular for $k=0.25,$ which jumps from
negligible values in Fig.~\ref{short-time-1} to
crossing the $k=10$ curve of the linear entropy in Fig.~%
\ref{short-time-2}(b). For the first few periods this increase of
the linear entropy takes place in steps, reflecting the collision
of the radially localized electron wavepacket with the core at
each half-integer value of $t/T_{e}$ (the wavepacket spreads
radially after a few periods). These findings will be discussed in
Sec.~IV but we may note again that as found for the eigenstates
there is no simple relation between the global classical dynamical
regime and quantum entanglement generation.

\subsubsection{Long-time evolution}

\begin{figure}[tb]
\includegraphics[height=2.5in,width=2.15in]{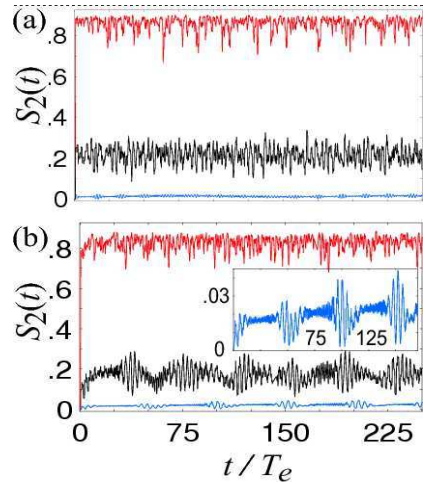}
\caption{Long time variation of the linear entropy for the case
considered in Fig.~\ref{short-time-1} ($ N_{0}=40$, (a) generic and
(b) resonant cases). The inset in (b) blows up the vertical
scale for the $k=0.25$ curve to help visualize the oscillations, also visible
for $k=1$.}%
\label{long1}%
\end{figure}

\begin{figure}[tb]
\includegraphics[height=2.3in,width=2.15in]{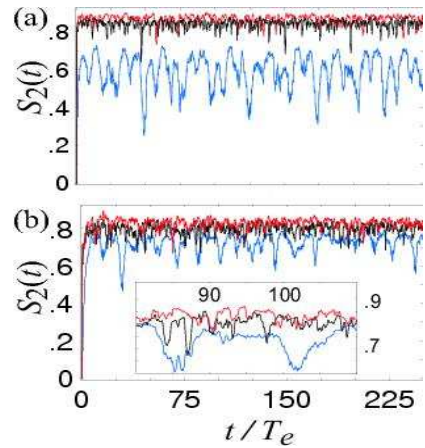}
\caption{Long time variation of the linear entropy for the case
considered in Fig.~\ref{short-time-2} ($ N_{0}=50$, (a) generic and
(b) resonant cases). The inset in (b) details $S_2(t)$ in the
region $t=$ $80$ to $110$ $T_e$. }%
\label{long2}%
\end{figure}

In Figs.~\ref{short-time-1} and \ref{short-time-2}, the linear
entropy appears to saturate after a few periods. For longer times
$S_{2}(t)$ is plotted in Figs.~\ref{long1} (when the initial state
has $N_{0}=40$) and \ref{long2} ($N_{0}=50$). In most of the
cases (all the $k=10$ curves and the $k=1$ curves except in
Fig.~\ref{long1}(b)) $S_{2}(t)$ appears to vary randomly around
some average value. However for small kicks (the $k=0.25$ curves,
but also the $k=1$ curve in Fig.~\ref{long1}(b)) the repetition
of oscillatory structures is clearly visible to the eye.
These repetitions are due to partial wavepacket
revivals within each channel, which take place when the terms that
control the spreading of the packet regain an approximate phase
coherence. The revival times within each channel are determined in
the semiclassical approximation by expanding the energies in the
exponentials in Eq.~(\ref{e37}) as a function of the classical
action. The core's revival time $T_{c}^{rev}\propto B_{r}^{-1}$ is
independent of the energy whereas $T_{e}^{rev}\propto\epsilon^{-2}$.
A revival in the linear entropy will be visible provided 1 or 2
channels dominate in the overall contribution.

\begin{figure}[tb]
\includegraphics[height=1.5in,width=2.5in]{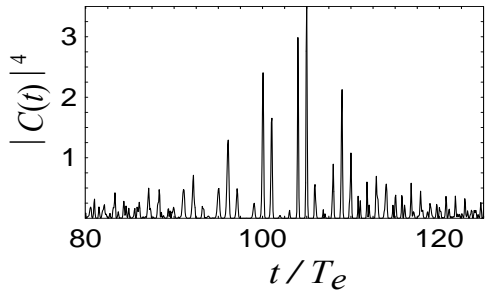}
\caption{The partial autocorrelation $C(t)$ defined by Eq.~%
(\ref{e91}) (parameters corresponding to the $k=0.25$ resonant
case, arbitrary units) is plotted in an interval centered around
the revival time
$T_{e}^{rev}\simeq 105$ $T_{e}$. }%
\label{long-corr}%
\end{figure}

For the $k=0.25$ case in Fig.~\ref{long1}(b), most of the
probability density stays in the original $N_0=40$ channel, a small
flux flowing to the $N=42$ channel. This flux is responsible for the
main part of the entanglement generation. We have plotted in
Fig.~\ref{long-corr} the correlation function
\begin{equation}
C(t)=\left\langle \psi_{42}(t=T_{e}/2)\right|  \left.  \psi_{42}%
(t)\right\rangle \label{e91}
\end{equation}
where%
\begin{equation}
\left|  \psi_{42}(t)\right\rangle \equiv\left\langle N=42\right|
\left. \psi(t)\right\rangle
\end{equation}
gives the electronic wavepacket in the $N=42$ channel. It can be
seen that near $t=100$ $T_e$ the correlation function oscillates
dramatically: high peaks appear while at the same time the lowest
values are near zero. This behavior translates into the
entanglement rate, which shows the same strong oscillations as
seen in the inset in Fig.~\ref{long1}. The revival time for the
electron motion in this case is computed in the semiclassical
approximation as $T_{e}^{rev}\simeq 105$ $T_{e}$. This value fits
well with the periodic repetitions of these strong oscillations
observed for $S_2(t)$.

\section{Discussion}

If we compare the time evolution of the linear entropy portrayed in
Fig.~\ref{short-time-1} to those of Fig.~\ref{short-time-2}, the importance
of the initial state is clear. If we further contrast these
results with the Poincar\'{e} surfaces of section of Fig.~\ref{figPSS} and
the localization of the initial states shown in Fig.~\ref{figPSS0max}, we
see that the generation of entanglement does not essentially depend on
the global classical dynamics, but rather on the specific dynamics
that leads to inelastic scattering. Classically, inelastic
scattering means that the value of $N$ changes during the
collision; then on the surface of section two consecutive points
cannot lie on circles around the $X$ axis (since these circles precisely
correspond to a fixed value of $N$). Inelastic collisions do take place
when the dynamic is chaotic: for very strong kicks ($k=10$)
two consecutive points on the Poincar\'{e} sections are arbitrarily
separated on the sphere.
Quantum mechanically inelastic scattering is translated into
superpositions of states having a different value of $N$. The
$k=10$ curves in Figs.~\ref{short-time-1}(a) and
\ref{short-time-2}(a) indeed reflect large and fast entanglement generation.
$S_{2}(t)$ achieves its maximal value (of 10/11) just after a couple
of collisions.

However inelastic scattering can also be induced by regular
dynamics. Consider $k=0.25$ in the resonant case. When the initial
state encircles the $X$ axis at the front of the sphere as happens
with $N_{0}=40$ (see the shaded region in Fig.~\ref{figPSS0max}), it will be
hardly modified by the classical dynamics. The lines in the
surface of section (Fig.~\ref{figPSS}(d)) also encircle the $X$ axis, thereby
conserving $N_{0}$: there is essentially only elastic scattering.
Quantum mechanically we expect little or no entanglement, as is
observed in Fig.~\ref{short-time-1}. But for $N_{0}=50$ the
initial distribution spans across the lines of regularity in the
surface of section, which are organized around the elliptic fixed
point on the $Z$ axis. These lines thus define torii in
phase-space that cut across several values of $N$, meaning
inelastic scattering and quantum-mechanical superpositions. Note
that in the generic $k=0.25$ situation the linear entropy is lower
than in the resonant case; classically the structure of phase
space in the zone covered by the initial state is modified due to
the appearance of resonant islands (Fig.~\ref{figPSS}(a)), leading to some
lines that roughly encircle the $X$ axis, partially favoring
elastic scattering.

We thus see that the entanglement production reflects the
classical dynamics in that the relevant parts of space-phase
leading to inelastic collisions (yielding quantum superpositions)
are explored with a large amplitude. To a large extent this
phenomenon is unrelated to whether the dynamical regime is chaotic
or regular. Classical chaos is not necessary to induce high
entanglement generation. In our system chaos is sufficient,
because it always leads to inelastic scattering. Hence eigenstates
quantized in an underlying strongly chaotic phase-space will
present on average a large amount of mixtures, as seen in Fig.~\ref{fig3}.
These remarks are in line with similar conclusions
\cite{seligman gorin03,kus04,fujisaki04} which contrarily to
earlier results \cite{furuya etal98,miller sarkar99}, do not
attribute to chaos a higher entangling power.

Let us mention that more precise investigations of the
correlations between classical dynamics and dynamically induced
entanglement should take into account more information than what
can be inferred from the Poincar\'{e} surfaces of sections. For
example in the generic $k=1$ case (Fig.~\ref{figPSS}(b)) diffusion in the
chaotic sea takes place at a considerably lower rate than in the
$k=10$ situation. In particular, the particle may be trapped for
several periods in certain regions of phase-space, diffusing
slowly in the relevant regions of the surface of section. This
type of phenomena has an influence on the quantization process,
and if important will influence the entanglement production, as in
the present case. We therefore expect that these system specific
features, along with the r\^{o}le of the initial state relative
to the precise structure of phase-space, may severely constrain
the applicability to realistic systems of general 'universal'
formulas ruling the entanglement generation in chaotic and regular
systems that have been recently obtained
\cite{jacquod04,prosen05}. Semiclassical universal formulas, based
on the global average properties of classical phase-space, are
important as they set the trend that is followed by a quantum
system with simple dynamics (like a uniform transition to chaos).
However individual features of the system are known to be
important in the semiclassical description of diffractive effects
induced by a coupling potential and need to be taken into account
e.g. to describe the spectral statistics \cite{mm04}. It is thus
not surprising to see that dynamical entanglement induced by a
standard potential coupling is correlated with the local structure
of phase-space, and not only with its global properties.

Concerning the long-time behavior, it may be noted that the revivals
appear well beyond the Heisenberg time, time after which quantum
phenomena having no classical counterpart become prominent. In some
cases, for example when the number of channels is small, the
oscillations due to wavepacket revivals induce very large variations
of the entanglement rate totally unrelated to the behavior at
short-times. Hence the linear entropy of a state which initially
only showed a slow and small amount of entanglement generation can
rise above the linear entropy of a state that initially displayed
high and fast entanglement production.

We finally point out that entanglement in Rydberg molecules is
routinely  detected experimentally, given that the consequences of
entanglement appear in even the most elementary measurements (e.g.
interference of Rydberg series in photoabsorption spectra). A
quantitative measurement of entanglement, that would more closely
reflect the evolution of the linear entropy, can be set up by
combining the methods employed for the detection of interfering
Rydberg wavepackets \cite{exp}. These methods are based on the use
of several laser pulses with well-defined phase relations to monitor
the interferences appearing in the population of predefined Rydberg
states.

\section{Conclusions}

We have investigated the entanglement production in Rydberg molecules and
contrasted the results with the underlying classical dynamics, which is
known to play a r\^{o}le in the understanding of observable spectroscopic %
 effects  \cite{lombardi88} and in the interpretation of the energy
levels statistics \cite{lombardi93}. We have first determined the
average linear entropy of eigenstates corresponding to different
collision strengths (quantum phase shifts or classical kicks) and
then studied the generation of entanglement from initially
unentangled states. We have seen that the quantum/classical
correspondence on the level of entanglement production is relevant
not on the scale of the global classical dynamical regime, but
rather on the specific classical features that
quantum-mechanically translate into superpositions. In Rydberg
molecules, it is a high rate of inelastic scattering for an
initial classical distribution that corresponds in the quantum
domain to states displaying a high degree of entanglement. The
relation between global chaotic or regular behavior on the one
hand and these specific classical features that will be translated
quantum mechanically as entanglement production on the other hand
appears to highly depend on the individual system under
investigation. We therefore conclude it seems unlikely that the
generation of entanglement could be employed as a reliable
signature of chaos for an arbitrary dynamical system.

\end{document}